\documentclass[aps,prb,preprint,amsmath,amssymb,floatfix]{revtex4}
\usepackage{amsmath}
\usepackage{amsfonts}
\usepackage{graphicx}
\newcommand{\id}{I}
\newcommand{\idn}{\mathbf 1}
\DeclareMathOperator{\Tr}{Tr}

\newcommand{\real}{\mathbf{R}}
\newcommand{\compl}{\mathbf{C}}

\newcommand{\cc}[1]{\overline{#1}}
\newcommand{\ve}[1]{\mathbf{#1}}
\newcommand{\vs}[1]{\boldsymbol{#1}}
\newcommand{\sr}[1]{\mathbf{#1}{\boldsymbol{\sigma}}}
\newcommand{\traf}[1]{#1'}
\newcommand{\traff}[1]{\tilde{#1}}

\newcommand{\conc}{\mathcal{C}}
\newcommand{\negat}{\mathcal{N}}

\begin{document}\
\title{A study of two-qubit density matrices with fermionic
purifications}
\author{Szil\'ard Szalay, P\'eter L\'evay, Szilvia Nagy and
J\'anos Pipek}

\affiliation{Department of Theoretical Physics, Institute of
Physics, Budapest University of Technology and Economics, H-1111
Budafoki \'ut 8, Hungary}
\date{\today}
\begin{abstract}
We study $12$ parameter families of two qubit density matrices,
arising from a special class of two-fermion systems with four
single particle states or alternatively from a four-qubit state
with amplitudes arranged in an antisymmetric matrix.  We calculate
the Wooters concurrences and the negativities in a closed form and
study their behavior. We use these results to show 
that the relevant entanglement measures satisfy 
the generalized Coffman-Kundu-Wootters formula of distributed
entanglement. An explicit formula for the residual tangle is also given.
The geometry of such density matrices is
elaborated in some detail. In particular an explicit form for the
Bures metric is given.

\end{abstract}

\maketitle{}

\section{Introduction}
Entanglement is the basic resource of quantum information
processing\cite{Nielsen}. As such it has to be quantified and its
structure characterized. For entanglement quantification one uses
special classes of entanglement measures which are real-valued
functions on the states. Pure and mixed state entanglement and its
quantification in its bipartite form is a 
well-understood phenomenon. Moreover, on the geometry and
structure of entangled states associated with such systems a large
number of interesting results is available\cite{Geom}.

 For example for pure states of
bipartite systems the classification of different entanglement
types is effected by the Schmidt decomposition. If the Schmidt
decomposition is known, from the Schmidt numbers one can form the
von-Neumann entropy\cite{Petz} as a good measure characterizing
bipartite entanglement. For quantifying mixed state entanglement
no such general method exists. For the special case of two qubits
as a measure of entanglement we have the celebrated formula of
Hill and Wootters\cite{Hill} for the bipartite {\it concurrence}
${\cal C}$ and the associated entanglement of formation. The
structure of this measure of entanglement was studied in many
different papers\cite{hivatbengtsson}. Its structure has been
related to antilinear operators\cite{Uhlmann}, combs and
filters\cite{Siewert}, and has also been generalized to
rebits\cite{Fuchs}. Explicit expressions for different special
classes of density matrices and a comparison with other measures
of entanglement has been given\cite{Eisert, Verstraete, Ishizaka}.

In this paper we would like to study the structure of special
$12$ parameter families of two-qubit density matrices for which the
the mixed state concurrences can be calculated in a closed form. Such
density matrices can be regarded as reduced ones coming from some
larger system with special properties. In order to motivate our
choice for this larger system we consider an example. If we
consider a three-qubit state $\vert\psi\rangle\in \compl^2\otimes
\compl^2\otimes \compl^2$, then after calculating any of the
reduced density matrices e.g. ${\varrho}_{12}={\rm
Tr}_{3}(\vert\psi\rangle\langle\psi\vert)$ we are left with a
two-qubit density matrix of very special structure. For example in
this case ${\varrho}_{12}$ is of rank two, and this observation
enables an explicit calculation of the mixed state concurrence in
terms of the amplitudes of the three-qubit state
$\vert\psi\rangle$. This result forms the basis of further
important developments namely the derivation of the
Coffman-Kundu-Wootters relation of distributed
entanglement\cite{Kundu}.

Proceeding by analogy we expect that four-qubit states of special
structure might provide us with further interesting examples of
that kind. Let us consider a four-qubit state $\vert\Psi\rangle\in
\compl^2\otimes \compl^2\otimes \compl^2\otimes \compl^2$. A
class of two-qubit density matrices arises after forming the
reduced density matrices like ${\varrho}_{12}={\rm
Tr}_{34}(\vert\Psi\rangle\langle\Psi\vert)$. However, density
matrices of that kind are still too general to have a
characteristic structure. Hence as an extra constraint we impose
an antisymmetry condition on the amplitudes of
\begin{equation}
\vert\Psi\rangle=\sum_{ijkl=0}^1{\Psi}_{ijkl}\vert ijkl\rangle,
\label{4}
\end{equation}
\noindent as
\begin{equation}
{\Psi}_{ijkl}=-{\Psi}_{klij}, \label{anti}
\end{equation}
\noindent i.e. we impose antisymmetry in the first and second {\it
pairs} of indices.

An alternative (and more physical) way is the one of imposing such
constraints on the original Hilbert space $H\simeq \compl^{16}$
which renders to have a tensor product structure on one of its six
dimensional subspaces ${\cal H}$ of the form
\begin{equation}
{\cal H}=(\compl^2\otimes \compl^2)\wedge (\compl^2\otimes \compl^2),
 \label{tps}
\end{equation}
\noindent where $\wedge$ refers to antisymmetrization. As we know
quantum tensor product structures are
observable-induced\cite{Zanardi}, hence in order to specify our
system with a tensor product structure of Eq.~(\ref{tps}) we have
to specify the experimentally accessible interactions and
measurements that account for the admissible operations we can
perform on our system. For example we can realize our system as a
pair of fermions with four single particle states where a part of
the admissible operations are local unitary transformations of
the form
\begin{equation}
\vert\Psi\rangle\mapsto (U\otimes V)\otimes (U\otimes
V)\vert\Psi\rangle,\qquad U,V\in U(2),\qquad \vert\Psi\rangle\in
{\cal H}. \label{local}
\end{equation}
\noindent Taken together with Eq.~(\ref{anti}) this transformation
rule clearly indicates that the first and the second and the third
and fourth subsystems form two {\it indistinguishable} subsystems
of fermionic type.

The aim of the present paper is to study the interesting structure
of the reduced density matrices of the form
${\varrho}_{ij}, \quad i<j, \quad i,j=1,2,3,4$ arising from
fermionic states that are elements of the tensor product structure
as shown by Eq.~(\ref{tps}). We can alternatively coin the term
that these density matrices are ones with {\it fermionic
purifications}.

The organization of the paper is as follows. In Section II. we
present our parametrized family of density matrices we wish to
study. Using suitable local unitary transformations we transform
this family to a canonical form. In Section III. based on these
results we calculate the Wootters concurrence the negativity and
the purity. We give a formula for the upper bound of negativity 
for a given concurrence. (We prove it in Appendix A.)
In section IV. we analyze the structure of these
quantities and discuss how they are related to each other.
In particular we prove that the relevant entanglement measures
associated with our four-qubit state satisfy the generalized Coffman-Kundu-Wootters inequality of distributed entanglement\cite{Osborne}.
For the residual tangle we derive an explicit formula, containing two from the four algebraically independent four-qubit invariants.
In
Section V. we investigate the Bures geometry of this special
subclass of two-qubit density matrices. We show that thanks to our
purifications being fermionic an explicit formula for the Bures
metric with hyperbolic structure can be obtained. The conclusions and some comments are left
for Section VI.

\section{The density matrix}

Let us parametrize the $6$ amplitudes of our normalized four qubit state
$\vert\Psi\rangle$ of Eq.~(\ref{4}) with the antisymmetry property
of Eq.~(\ref{anti}) as
\begin{equation}
\Psi_{ijkl}=\frac{1}{2}\left({\varepsilon}_{ik}{\cal A}_{jl}+{\cal
B}_{ik}{\varepsilon}_{jl}\right), \label{4para}
\end{equation}
\noindent where ${\cal A}$ and ${\cal B}$ are {\it symmetric}
matrices of the form
\begin{equation}
{\cal A} = \begin{pmatrix}
z_1-iz_2&-z_3\\
-z_3&-z_1-iz_2
\end{pmatrix}= 
\varepsilon({\bf z}\overline{\boldsymbol{\sigma}}), \quad
{\cal B}=\begin{pmatrix}
w_1-iw_2&-w_3\\
-w_3&-w_1-iw_2
\end{pmatrix}=
\varepsilon({\bf w}\overline{\boldsymbol{\sigma}}), \label{szimm}
\end{equation}
\noindent where ${\bf z},{\bf w}\in \compl^3$,
$\sr{w}= w_1\sigma_1 + w_2\sigma_2 + w_3\sigma_3$,
with the usual $\sigma_i$ Pauli matrices,
\begin{equation}
\varepsilon=\begin{pmatrix}0&1\\-1&0\end{pmatrix} \label{vareps}
\end{equation}
\noindent 
and the overline refers to complex conjugation.

It is straightforward to check, that the normalization condition
of the state $\vert\Psi\rangle$ takes the form:
\begin{equation}
\label{eq_psinorm}
 \Vert \Psi \Vert^2 \equiv \Vert \ve{w} \Vert^2 + \Vert \ve{z} \Vert^2 := 1
\end{equation}
The density matrices we wish to study are arising as
reduced ones of the form
\begin{equation}
\varrho\equiv \varrho_{12}={\rm
Tr}_{34}(\vert\Psi\rangle\langle\Psi\vert). \label{densi}
\end{equation}
\noindent Notice that since the $(12)$ and $(34)$ subsystems are
by definition {\it indistinguishable} we also have
$\varrho={\varrho}_{12}=\varrho_{34}$.

A calculation of the trace yields the following explicit form for
$\varrho$
\begin{equation}
 \label{eq_rho}
 \varrho=\frac{1}{4}\left( \idn + \Lambda \right),
\end{equation}
where $\idn$ denotes the $4\times 4$ identity matrix, and
$\Lambda$ is the traceless matrix
\begin{gather}
\label{eq_Lambda}
 \Lambda := \sr{x}\otimes\id + \id\otimes\sr{y}
+\sr{w}\otimes\sr{\cc{z}} +\sr{\cc{w}}\otimes\sr{z},\\
\label{eq_xy}
 \ve{x} := i\ve{w}\times\ve{\cc{w}}, \qquad
 \ve{y} := i\ve{z}\times\ve{\cc{z}},
\end{gather}
where $\id$ is the $2\times 2$ identity matrix. 
Notice, that $\ve{x}, \ve{y}
\in \real^3$, and
$\ve{x}\ve{w}=\ve{x}\ve{\cc{w}}=\ve{y}\ve{z}=\ve{y}\ve{\cc{z}}=0$.
Due to this, and the identities
\begin{equation}
\label{eq_xynorm}
 \vert \ve{x} \vert^2 = \Vert \ve{w} \Vert^4 - \ve{w}^2\ve{\cc{w}}^2, \qquad
 \vert \ve{y} \vert^2 = \Vert \ve{z} \Vert^4 - \ve{z}^2\ve{\cc{z}}^2,
\end{equation}
it can be checked, that $\Lambda$ satisfies the identity
\begin{equation}
 \label{eq_Lambda2}
 \Lambda^2 = (1-\eta^2) \idn,
\end{equation}
where
\begin{gather}
\label{eq_ceta}
 \eta \equiv\vert \ve{w}^2-\ve{z}^2 \vert, \\ 
\label{eq_etarng}
 0 \leq \eta \leq 1,
\end{gather}
Notice that the quantity
$\eta$ is just the Schliemann-measure of entanglement for
two-fermion systems with $4$ single particle
states\cite{Schlie,LNP}. Indeed our density matrix $\varrho$ (with
a somewhat different parametrization) can alternatively be
obtained\cite{LNP} as a reduced one arising from such fermionic
systems after a convenient global $U(4)$, and a local $U(2)\times
U(2)$  transformation of the form $I\otimes {\sigma}_2$.

Now by employing suitable local unitary transformations we would
like to obtain a canonical form for $\varrho$. According to 
Eq.~(\ref{local}) the transformations operating on subsystems $12$ or
equivalently $34$ are of the form $U\otimes V\in U(2)\times U(2)$.

As a first step let us consider the unitary transformation
\begin{equation}
\label{eq_tafU}
 U(\ve{\hat{u}}, \alpha):= e^{i\frac{\alpha}{2}\sr{\hat{u}}}=
 \cos\left(\frac{\alpha}{2}\right)\id + i\sin\left(\frac{\alpha}{2}\right)\sr{\hat{u}}.
\end{equation}
which is a spin-$\frac{1}{2}$ representation of an $SU(2)$
rotation around the axis $\ve{\hat{u}} \in \real^3$, ($\vert
\ve{\hat{u}} \vert = 1$) with an angle $\alpha$. A special
rotation from $\ve{x}$ to $\ve{\traf{x}}$ (${\bf x}^{\prime}\neq
-{\bf x}$) can be written as
\begin{gather}
 U(\ve{\hat{u}}, \alpha)^\dagger (\sr{x}) U(\ve{\hat{u}}, \alpha) = \sr{\traf{x}},\\
\label{eq_specU}
 U(\ve{\hat{u}}, \alpha) = \frac{1}{\sqrt{2\ve{x}^2(\ve{x}^2 + \ve{x}\ve{\traf{x}})}}
\left(\ve{x}^2 \id + (\sr{x})(\sr{\traf{x}})\right),\\
 \ve{\hat{u}} = \frac{\ve{x}\times \ve{\traf{x}}}{\vert\ve{x}\times \ve{\traf{x}}\vert},\qquad
 \alpha = \arccos(\frac{\ve{x}\ve{\traf{x}}}{\ve{x}\ve{x}} ).
\end{gather}

Employing this, we can rotate the vector $\ve x$ to the direction of the
coordinate axis $z$.
In this case
\begin{gather}
\label{eq_r}
r := \vert \ve{x}\vert,\\
U_\ve{x}:=\frac{1}{\sqrt{2r(r+x_3)}}\left(r\id + (\sr{x})\sigma_3\right),\\
U^\dagger_\ve{x}(\sr{x})U_\ve{x} = r\sigma_3.
\end{gather}
Moreover, using Eq.~(\ref{eq_specU}) it can be checked that due to the
special form of $\ve{x}$ (see Eq.~(\ref{eq_xy})), the
transformation above rotates the third component of $\ve w$ into
zero
\begin{gather}
 U^\dagger_\ve{x}(\sr{w})U_\ve{x} = \sr{\traf{w}},\\
\label{eq_trw}
\ve{\traf{w}} = \ve{w} - \frac{\ve{w}\ve{\traf{x}}}{r^2+\ve{x}\ve{\traf{x}}}(\ve{x} + \ve{\traf{x}})
 = \begin{bmatrix}
w_1-\frac{x_1}{r+x_3}w_3\\
w_2-\frac{x_2}{r+x_3}w_3\\
0
\end{bmatrix}.
\end{gather}
A similar set of transformations can be applied to $\sr{y}$
\begin{gather}
\label{eq_s}
s:=\vert \ve{y}\vert,\\
V_\ve{y}:=\frac{1}{\sqrt{2s(s+y_3)}}\left(s\id + (\sr{y})\sigma_3\right),\\
V^\dagger_\ve{y}(\sr{y})V_\ve{y} = s\sigma_3.\\
 V^\dagger_\ve{y}(\sr{z})V_\ve{y} = \sr{\traf{z}},\\
\label{eq_trz}
 \ve{\traf{z}} = \ve{z} - \frac{\ve{z}\ve{\traf{y}}}{s^2+\ve{y}\ve{\traf{y}}}(\ve{y} + \ve{\traf{y}})
 = \begin{bmatrix}
z_1-\frac{y_1}{s+y_3}z_3\\
z_2-\frac{y_2}{s+y_3}z_3\\
0
\end{bmatrix}.
\end{gather}

Obviously, every $U \in U(2)$ unitary transformation acting on an
arbitrary $\ve{a}\in \compl^3$ as $U^\dagger \sr{\ve{a}} U =
\sr{\ve{\traf{a}}} $ preserves $\ve{a}^2$ and $\Vert \ve{a}
\Vert^2$, since $\ve{a}^2 = -\det(\sr{\ve{a}})$, and $\Vert \ve{a}
\Vert^2 =
 \frac{1}{2}\Tr((\sr{\ve{a}})^\dagger(\sr{\ve{a}}))$.
Hence
\begin{gather}
\label{eq_wz2invar}
 \ve{\traf{w}}^2 = \ve{w}^2, \qquad
 \ve{\traf{z}}^2 = \ve{z}^2,\\
\label{eq_wza2invar}
 \Vert\ve{\traf{w}}\Vert^2 = \Vert\ve{w}\Vert^2, \qquad
 \Vert\ve{\traf{z}}\Vert^2 = \Vert\ve{z}\Vert^2.
\end{gather}
and
\begin{equation}
 \traf{\eta} = \eta
\end{equation}
are invariant under local $U(2)\times U(2)$ transformations. (The
entanglement measure  $\eta$ is also invariant under the larger
group of $U(4)$ transformations.)

Now by employing the local $U(2)\times U(2)$ transformations
$U_{\bf x}\otimes V_{\bf y}$, our density matrix can be cast to
the form,
\begin{gather}
\label{eq_rhotraf}
\traf{\varrho} =
\left(U_\ve{x}\otimes V_\ve{y}\right)^\dagger \varrho \left(U_\ve{x}\otimes V_\ve{y}\right)
= \frac{1}{4}\left(\idn + \traf{\Lambda} \right),\\
\label{eq_Lamtraf}
 \traf{\Lambda} = r\sigma_3\otimes\id + \id\otimes s\sigma_3
+\sr{\traf{w}}\otimes\sr{\traf{\cc{z}}} +\sr{\traf{\cc{w}}}\otimes\sr{\traf{z}},
\end{gather}
where $\traf{\Lambda}$  has the  special form
\begin{equation}
\label{eq_Xshape}
 \traf{\Lambda} = \begin{bmatrix}
\alpha_3 & 0 & 0 & \alpha_1-i\alpha_2\\
0 & \beta_3 & \beta_1-i\beta_2 & 0\\
0 & \beta_1+i\beta_2 & -\beta_3 & 0\\
\alpha_1+i\alpha_2 & 0 & 0 & -\alpha_3
\end{bmatrix}
\end{equation}
with the quantities $\vs{\alpha}, \vs{\beta} \in \real^3$ defined
as
\begin{align}
\label{eq_ab}
\vs{\alpha} &= \begin{bmatrix}
 \xi_1 - \xi_2 \\
 \zeta_1 + \zeta_2 \\
 r + s
\end{bmatrix},& \quad
\vs{\beta}  &= \begin{bmatrix}
 \xi_1 + \xi_2 \\
 \zeta_1 - \zeta_2 \\
 r - s
\end{bmatrix},&\\
\label{eq_abe}
\xi_1 &= \traf{w}_1 \traf{\cc{z}}_1 + \traf{\cc{w}}_1 \traf{z}_1,& \quad
\zeta_1 &= \traf{w}_2 \traf{\cc{z}}_1 + \traf{\cc{w}}_2 \traf{z}_1, \\
\label{eq_abv}
\xi_2 &= \traf{w}_2 \traf{\cc{z}}_2 + \traf{\cc{w}}_2 \traf{z}_2,& \quad
\zeta_2 &= \traf{w}_1 \traf{\cc{z}}_2 + \traf{\cc{w}}_1 \traf{z}_2.
\end{align}
Thanks to the special shape of ${\Lambda}$, we can regard
$\traf{\varrho}$ as the direct sum of two $2\times 2$ blocks, i.e.
$(\id+\sr{\vs{\alpha}})$ and $(\id+\sr{\vs{\beta}})$. Having
obtained the canonical form of our reduced density matrix
$\varrho$, now we turn to the calculation of the corresponding
entanglement measures.

\section{Measures of entanglement for the density matrix}
\subsection{Concurrence}
\label{sec_Concurr}

In this section we calculate the Wootters-concurrence \cite{Hill}
of our density matrix $\varrho$ defined in Eqs.~(\ref{eq_rho}) -
(\ref{eq_xy}). This quantity is defined as
\begin{equation}
{\cal C}={\rm max}\{0,\lambda_1-\lambda_2-\lambda_3-\lambda_4\}
\label{eigen}
\end{equation}
\noindent where $\lambda_1\geq\lambda_2\geq\lambda_3\geq\lambda_4$
are the {\it square roots} of the eigenvalues of the
matrix\cite{Hill} $\varrho\tilde{\varrho}$ where

\begin{equation}
\label{eq_sf}
 \tilde{{\varrho}} = (\sigma_2 \otimes \sigma_2){\overline{\varrho}}
  (\sigma_2 \otimes \sigma_2).
\end{equation}
\noindent This matrix (the Wootters spin-flip of $\varrho$) is
known to have real nonnegative eigenvalues. Moreover, the
important point is that ${\cal C}$ is an $SL(2,\compl)\times
SL(2, \compl)$ invariant\cite{Siewert}, hence we can use the
canonical form we obtained in the previous section via using the
transformation $U_{\bf x}\otimes V_{\bf y}\in SU(2)\times
SU(2)\subset SL(2, \compl)\times SL(2, \compl)$ for its
calculation.

It is straightforward to check that  $16
\traf{\varrho}\tilde{\traf{\varrho}} $ has the same X-shape as
$\traf{\varrho}$, with the blocks
$(\traff{\alpha}_0\id+\sr{\traff{\vs{\alpha}}}) $ and
$(\traff{\beta}_0\id+\sr{\traff{\vs{\beta}}})$
where $\traff{\alpha}_\mu$, $\traff{\beta}_\nu$ $\mu,\nu=0,1,2,3$
are quadratic in $\vs{\alpha}$ $\vs{\beta}$:
\begin{equation}
\label{eq_abtraff}
\begin{split}
\traff{\alpha}_\mu =
\begin{bmatrix}
1+\alpha_1^2+\alpha_2^2-\alpha_3^2\\
2\alpha_1 -i2\alpha_2\alpha_3\\
2\alpha_2 +i2\alpha_3\alpha_1\\
0
\end{bmatrix} \in \compl^4,\\
\traff{\beta}_\nu =
\begin{bmatrix}
1+\beta_1^2+\beta_2^2-\beta_3^2\\
2\beta_1 -i2\beta_2\beta_3\\
2\beta_2 +i2\beta_3\beta_1\\
0
\end{bmatrix} \in \compl^4.
\end{split}
\end{equation}
The eigenvalues of the blocks 
$(\traff{\alpha}_0\id+\sr{\traff{\vs{\alpha}}})$ and
$(\traff{\beta}_0\id+\sr{\traff{\vs{\beta}}})$ are
$\traff{\alpha}_0 \pm \sqrt{ \traff{\vs{\alpha}}^2 } $ and
$\traff{\beta}_0 \pm \sqrt{ \traff{\vs{\beta}}^2 } $,
respectively. Now, we can express these with the help of the
quantities $\vs{\alpha}$, $\vs{\beta}$ of (\ref{eq_abtraff}) and
get the eigenvalues of $ \traf{\varrho}\tilde{\traf{\varrho}}$ in
the form
\begin{equation}
\begin{split}
 \Lambda_i =
\Biggl\{
&\frac{1}{16}\left(\sqrt{\alpha_1^2+\alpha_2^2} \pm \sqrt{1-\alpha_3^2}\right)^2,\\
&\frac{1}{16}\left(\sqrt{\beta_1^2+\beta_2^2} \pm \sqrt{1-\beta_3^2}\right)^2
\Biggr\}
\end{split}
\end{equation}
Now, using Eqs.~(\ref{eq_ab}) - (\ref{eq_abv}), we have to express these
as functions of our original quantities $\ve{z}^2$, $\ve{w}^2$,
$\Vert\ve{z}\Vert^2$ and $\Vert\ve{w}\Vert^2$. Straightforward
calculation shows, that:
\begin{equation}
\begin{split}
\label{eq_jajj}
\alpha_1^2+\alpha_2^2 &=
2\Vert\traf{\ve{w}}\Vert^2\Vert\traf{\ve{z}}\Vert^2 +
\traf{\ve{w}}^2\traf{\cc{\ve{z}}}^2 + \traf{\cc{\ve{w}}}^2\traf{\ve{z}}^2-2rs,\\
\beta_1^2+\beta_2^2 &=
2\Vert\traf{\ve{w}}\Vert^2\Vert\traf{\ve{z}}\Vert^2 +
\traf{\ve{w}}^2\traf{\cc{\ve{z}}}^2 + \traf{\cc{\ve{w}}}^2\traf{\ve{z}}^2+2rs,\\
1-\alpha_3^2 &=
2\Vert\traf{\ve{w}}\Vert^2\Vert\traf{\ve{z}}\Vert^2 +
\traf{\ve{w}}^2\traf{\cc{\ve{w}}}^2 + \traf{\ve{z}}^2\traf{\cc{\ve{z}}}^2-2rs,\\
1-\beta_3^2 &=
2\Vert\traf{\ve{w}}\Vert^2\Vert\traf{\ve{z}}\Vert^2 +
\traf{\ve{w}}^2\traf{\cc{\ve{w}}}^2 + \traf{\ve{z}}^2\traf{\cc{\ve{z}}}^2+2rs.
\end{split}
\end{equation}
The formulas above are expressed in terms of quantities invariant
under our transformation yielding the canonical form (see 
Eqs.~(\ref{eq_wz2invar})-(\ref{eq_wza2invar})), hence we can simply
omit the primes. Hence by using Eq.~(\ref{eq_xynorm}) and
(\ref{eq_ceta}) we can establish that
\begin{equation}
\begin{split}
\alpha_1^2+\alpha_2^2 &= 1-\alpha_3^2 - \eta^2, \\
\beta_1^2+\beta_2^2 &= 1-\beta_3^2 - \eta^2.
\end{split}
\end{equation}
For further use, denote:
\begin{equation}
\label{gammapm}
 \gamma_+ := r+s \equiv \alpha_3, \qquad
 \gamma_- := r-s \equiv \beta_3.
\end{equation}
With these, the square root of the eigenvalues of
$\varrho\tilde{\varrho}$ are
\begin{equation}
\begin{split}
\lambda_i = \sqrt{ \Lambda_i} =
\Biggl\{
&\frac{1}{4}\left( \sqrt{1-\gamma_+^2} \pm \sqrt{1-\gamma_+^2-\eta^2} \right),\\
&\frac{1}{4}\left( \sqrt{1-\gamma_-^2} \pm
\sqrt{1-\gamma_-^2-\eta^2} \right). \Biggr\}
\end{split}
\end{equation}
The biggest one of these is $\lambda_{max} = \frac{1}{4}\left(
\sqrt{1-\gamma_-^2} + \sqrt{1-\gamma_-^2-\eta^2} \right) $ and
after subtracting the others from it, we get finally the nice
formula for the concurrence
\begin{equation}
 \label{eq_conc}
 \conc (\varrho) = \max\left\{ 0,
\frac{1}{2}\left( \sqrt{1-\gamma_-^2-\eta^2}- \sqrt{1-\gamma_+^2}\right) \right\}
\end{equation}
with the quantities defined in Eqs.~(\ref{eq_xy}), (\ref{eq_ceta}),
(\ref{eq_r}), (\ref{eq_s}) and (\ref{gammapm}) containing our
basic parameters ${\bf w}$ and ${\bf z}$ of $\varrho$.
One can easily check by the definitions (\ref{gammapm}),
that the surface dividing the entangled and separable states
in the space of these density matrices
is a special deformation of the $\eta = 0$ Klein-quadric, \cite{LNP}
given by the equation:
\begin{equation}
\eta^2 = 4rs.
\end{equation}
This can be also seen from the (\ref{eq_negat}) formula of negativity,
see in next subsection.

\subsection{Negativity} \label{sec_Neg}

Another entanglement-measure which we can calculate for $\varrho$
is the negativity. It is related to the notion of partial
transpose and the criterion of Peres\cite{Peres}. It is defined by
the smallest eigenvalue of the partially transposed density
matrix, as follows\cite{Geom,Verstraete}
\begin{equation}
\label{eq_neg}
  \negat (\varrho) = \max \left\{ 0, -2 \mu_{min} \right\}.
\end{equation}
Since the eigenvalues of a complex $4\times 4$ matrix are
invariant under $U(4)$ transformation, we can use again the
$SU(2)\times SU(2) \subset U(4)$-transformed $\traf{\varrho}$ of
Eq.~(\ref{eq_rhotraf}).

Denote by $\traf{\varrho}^{T_2}$ the partial transpose of
$\traf{\varrho}$ with res\-pect to the second subsystem. This
operation results in the transformation $ \traf{\Lambda}^{T_2} =
r\sigma_3\otimes\id + \id\otimes s\sigma_3
+\sr{\traf{w}}\otimes(\sr{\traf{\cc{z}}})^T
+\sr{\traf{\cc{w}}}\otimes(\sr{\traf{z}})^T$ i.e. only
$\traf{z}_2$ changes to $-\traf{z}_2$. By virtue of this,
retaining the (\ref{eq_abe}), (\ref{eq_abv}) definitions of
$\xi_1$, $\xi_2$, $\zeta_1$, $\zeta_2$, and redefining
$\vs{\alpha},\vs{\beta}\in \real^3$ of Eq.~(\ref{eq_ab}) as
\begin{equation}
\label{eq_abre}
\vs{\alpha} = \begin{bmatrix}
 \xi_1 + \xi_2 \\
 \zeta_1 - \zeta_2 \\
 r + s
\end{bmatrix}, \quad
\vs{\beta}  = \begin{bmatrix}
 \xi_1 - \xi_2 \\
 \zeta_1 + \zeta_2 \\
 r - s
\end{bmatrix},
\end{equation}
the calculation proceeds as in section \ref{sec_Concurr}.
The eigenvalues of $\varrho^{T_2}$ are $\mu_i =
\{\frac{1}{4}(1 \pm \vert \vs{\alpha} \vert), \frac{1}{4}(1 \pm \vert \vs{\beta} \vert) \}$,
and straightforward calculation shows, that
$\vs{\alpha}^2= 1 - \eta^2 + 4rs$ and
$\vs{\beta}^2 = 1 - \eta^2 - 4rs$.
Hence one can see, that the negativity of $\varrho$ is
\begin{equation}
 \label{eq_negat}
 \negat (\varrho) = \max \left\{ 0,
    \frac{1}{2} \left( \sqrt{ 1 - \eta^2 + 4rs } -1 \right)  \right\},
\end{equation}
with the usual conventions of equations (\ref{eq_ceta}),
(\ref{eq_r}) and (\ref{eq_s}).

%

\subsection{Comparsion of concurrence and negativity} \label{sec_ConcNeg}
For a $2$-qubit density matrix
we can write the following inequalities between the concurrence and the negativity
\begin{equation}
\label{eq_Vineq}
 \sqrt{ (1-\conc)^2 + \conc^2 } - (1-\conc) 
\leq \negat \leq \conc,
\end{equation}
which are known from a paper of Audenaert et.~al.\cite{Verstraete}
Our special case with fermionic correlations may give extra restrictions between 
concurrence and negativity, so we can pose the question,
whether we can replace inequality (\ref{eq_Vineq}) by a stronger one.

\begin{figure}[!ht]
 \centering
 \includegraphics[width=0.5\textwidth]{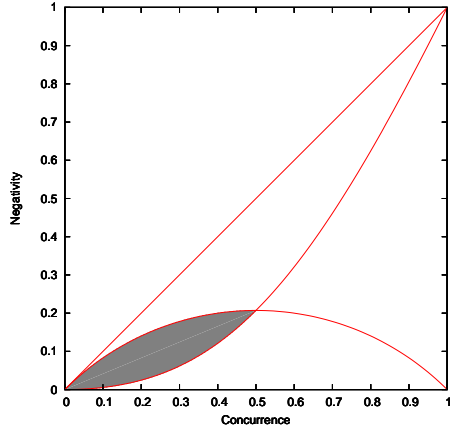}
 \caption{ Range of value of negativity for a given concurrence,
with its boundaries, as in (\ref{eq_ineq}). }
 \label{fig_ejjha}
\end{figure}

\textit{Fig.~\ref{fig_ejjha}} shows the result of a numerical calculation.
The gray field denotes the possible entangled states of the (\ref{eq_rho})-form
on the $\conc - \negat$ plane.
The upper bound of these can be analitically determined,
and it can be proven, (see Appendix \ref{sec_app}) that the following inequality holds
for $\negat$:
\begin{equation} \\
\label{eq_ineq}
\sqrt{ (1-\conc)^2 + \conc^2 } - (1-\conc) 
\leq \negat
\leq \frac{1}{2} \left( \sqrt{ 2 - ( 1-2\conc)^2 } -1 \right)
\leq \conc
\end{equation}
To see, that this upper bound is the tightest, consider
the special case, when $\ve{w} = \ve{z}$.
These states realize the boundary, so
the second inequality in (\ref{eq_ineq}) turns to equality.
(In this case $\eta = 0$, $r=s$, $\gamma_+= 2r$, $\gamma_-= 0$,
and for entangled states,
$\conc = \frac{1}{2}(1-\sqrt{1-4r^2})$,
$\negat = \frac{1}{2}(\sqrt{1+4r^2}-1)$.
These depend only on $r$, wich can be expressed from $\conc$,
thus we can express the negativity of these states with their concurrence,
and get back the curve of the upper bound.)

It can be seen, 
by calculating the intersection of the corresponding curves of (\ref{eq_ineq}),
that for maximally entangled states 
$\conc(\varrho_{max}) = \frac{1}{2}$, and
$\negat(\varrho_{max}) = \frac{ \sqrt{2} - 1 }{2}$.
We can study the behavior of these states: 
from the (\ref{eq_conc}) concurrence formula one can see, 
that
\begin{equation}
\conc = \frac{1}{2} \quad\Longleftrightarrow\quad
\left( \eta^2=0, \quad\text{and}\quad  \gamma_-^2=0, \quad\text{and} \quad \gamma_+^2=1 \right).
\end{equation}
The first two of these constraints hold, 
if and only if $\ve{w}^2=\ve{z}^2$, and $r=s$,
because of (\ref{eq_ceta}), (\ref{eq_xynorm}) and (\ref{gammapm}).
If $\ve{w}^2=\ve{z}^2$ then $r=s$ and $\Vert\ve{w}\Vert^2=\Vert\ve{z}\Vert^2 = \frac{1}{2}$
are equivalent,
and if $r=s$ then $\gamma_+^2=4r^2=1$,
$\Vert\ve{w}\Vert^4-\vert\ve{w}^2\vert^2=\frac 1 4$
and follows, that $\ve{w}^2=0$.
\begin{equation}
\label{concmax}
\conc = \conc_{max} = \frac{1}{2} \quad\Longleftrightarrow\quad 
\left( \Vert\ve{w}\Vert^2=\Vert\ve{z}\Vert^2 = \frac{1}{2},  \quad\text{and} \quad \ve{w}^2=\ve{z}^2 = 0 \right).
\end{equation}
Since the (\ref{eq_rhotraf}) transformation on $\varrho$
preserves the quantities appearing here, 
we can easily calculate the (\ref{eq_rhotraf}) canonical form
of the maximally entangled state $\traf{\varrho}$.
Let us choose an ansatz of the form (\ref{eq_trw}) and (\ref{eq_trz}) 
for $\traf{\ve{w}_{max}}$ and $\traf{\ve{z}_{max}}$
as $\traf{\ve{w}_{max}} = \frac{1}{\sqrt{2}} \begin{bmatrix} 
\cos(\alpha) e^{i\varphi_1}, \sin(\alpha) e^{i\varphi_2}, 0 \end{bmatrix}^T $
and $\traf{\ve{z}_{max}} = \frac{1}{\sqrt{2}} \begin{bmatrix} 
\cos(\beta) e^{i\psi_1}, \sin(\beta) e^{i\psi_2}, 0 \end{bmatrix}^T $.
These satisfy the first constraint of (\ref{concmax}),
and from the second follows that 
$\cos{\alpha}=\sin{\alpha}=\frac{1}{\sqrt{2}}$ and 
$\varphi_1 = \varphi_2-\frac{\pi}{2} =:\varphi$ and the same for $\traf{\ve{z}_{max}}$.
\begin{equation}
\traf{\ve{w}_{max}}=\frac{1}{2} e^{i\varphi}  \begin{bmatrix} 1\\i\\0 \end{bmatrix},
\qquad
\traf{\ve{z}_{max}}=\frac{1}{2} e^{i\psi}  \begin{bmatrix} 1\\i\\0 \end{bmatrix}.
\end{equation}
Then for the density matrix with maximal concurrence we get the expression
\begin{equation}
\traf{\varrho}_{max} = \frac{1}{4} \begin{bmatrix}
 \frac{3}{2}  & 0 & 0 & 0\\
  0 &  1 & e^{i\delta}  & 0\\
  0 & e^{-i\delta} & 1 & 0\\
  0 & 0 & 0 & \frac{1}{2}
  \end{bmatrix},
\end{equation}
with $\delta = \varphi-\psi $ is the only parameter characterizing this
maximally entangled density matrix.

\subsection{Purity} \label{sec_Pur} The purity is measuring the degree of mixedness of a density matrix.
For our $\varrho$ thanks to the special property of $\Lambda$
(see in Eq.~(\ref{eq_Lambda2})) it can easily be calculated. We have the result
\begin{gather}
 \Tr\varrho^2 = \frac{1}{4} (2-\eta^2),\\
 \frac{1}{4} \leq \Tr \varrho^2 \leq \frac{1}{2}
\end{gather}
by virtue of Eq.~(\ref{eq_etarng}). The participation ratio is by
definition
\begin{equation}
 R = \frac{1}{\Tr \varrho^2} = \frac{4}{2-\eta^2}.
\end{equation}

\section{Relating different measures of entanglement}

Now we would like to discuss the physical meaning of our
quantities derived in the previous section. First of all let us
notice that the
\begin{equation}
{\varrho}_1={\rm Tr}_{234}(\vert\Psi\rangle\langle\Psi\vert)={\rm
Tr_2}({\varrho}_{12})={\rm Tr}_2(\varrho)=\frac{1}{2}(I+{\bf
x}\boldsymbol{\sigma}) \label{red1}
\end{equation}
\noindent
\begin{equation}
\varrho_2={\rm Tr}_{134}(\vert\Psi\rangle\langle\Psi\vert)={\rm
Tr_1}(\varrho_{12})={\rm Tr}_1(\varrho)=\frac{1}{2}(I+{\bf
y}\boldsymbol{\sigma})\label{red2}
\end{equation}
\noindent  reduced density matrices describe the entanglement
properties of subsystems $1$ and $2$ to the rest of the system
described by the four-qubit state $\vert\Psi\rangle$. It is
well-known that the measures describing how much these subsystems
are entangled with the rest are ${\cal C}^2_{1(234)}=4{\rm
Det}(\varrho_1)$ and ${\cal C}^2_{2(134)}=4{\rm Det}(\varrho_2)$. Due
to Eqs.~(\ref{red1}-\ref{red2}) these quantities are
\begin{equation}
0\leq {\cal C}^2_{1(234)}=1-r^2\leq 1,\qquad
0\leq {\cal C}^2_{2(134)}=1-s^2\leq 1.
\label{1redmertek}
\end{equation}
\noindent 
Clearly, since ${\varrho}_1={\varrho}_3$ and ${\varrho}_2={\varrho}_4$, we also have
\begin{equation}
{\cal C}^2_{3(124)}={\cal C}^2_{1(234)},\qquad
{\cal C}^2_{4(123)}={\cal C}^2_{2(134)}.
\label{1concurrences}
\end{equation}
\noindent

Moreover, we already know that ${\varrho}_{12}={\varrho}_{34}={\varrho}$.
A straightforward calculation of the two-partite density matrices ${\varrho}_{14}$ and ${\varrho}_{23}$ shows that they again have the form of Eq. (10)
with the sign of ${\bf w}$ is changed in the first case and the vectors ${\bf w}$ and ${\bf z}$ are exchanged in the second. Since these transformations do not
change the value of the concurrence, we have
\begin{equation}
{\cal C}^2_{12}={\cal C}^2_{14}={\cal C}^2_{23}={\cal C}^2_{34}.
\label{2concurrences}
\end{equation}
\noindent
Now the only two-qubit density matrices we have not discussed yet are the ones
${\varrho}_{13}$ and ${\varrho}_{24}$.
Their form is
\begin{equation}
\left({\varrho}_{13}\right)_{iki^{\prime}k^{\prime}}=\frac{1}{2}\left(\vert\vert {\bf z}\vert\vert^2{\varepsilon}_{ik}{\varepsilon}_{i^{\prime}k^{\prime}}+
{\cal B}_{ik}\overline{\cal B}_{i^{\prime}k^{\prime}}\right),\qquad
\left({\varrho}_{24}\right)_{jlj^{\prime}l^{\prime}}=\frac{1}{2}\left(\vert\vert {\bf w}\vert\vert^2{\varepsilon}_{jl}{\varepsilon}_{j^{\prime}l^{\prime}}+
{\cal A}_{jl}\overline{\cal A}_{j^{\prime}l^{\prime}}\right).
\label{ujdensity}
\end{equation}
\noindent
Recall now the that the (\ref{local}) transformation property of our four-qubit state gives rise to the corresponding ones for the reduced density matrices
\begin{equation}
{\varrho}_{13}\mapsto (U\otimes U){\varrho}_{13}(\overline{U}\otimes \overline{U}),\qquad
{\varrho}_{24}\mapsto (V\otimes V){\varrho}_{24}(\overline{V}\otimes \overline{V}).
\label{transro}
\end{equation}
\noindent
For $U,V\in SU(2)$ we have $V{\varepsilon}V^t=U{\varepsilon}U^t={\varepsilon}$, hence the tensors occurring in Eq. (\ref{ujdensity}) transform as
\begin{equation}
{\varepsilon}\mapsto {\varepsilon},\qquad {\cal A}\mapsto V{\cal A}V^t,\qquad
{\cal B}\mapsto U{\cal B}U^t.
\label{trnsf}
\end{equation}
Using the (\ref{szimm}) definition of ${\cal A}$ 
we have for example
\begin{equation}
V{\cal A}V^t=\varepsilon\overline{V}{\bf z\overline{\boldsymbol{\sigma}}}V^t=\varepsilon\overline{V\overline{\bf z}{\boldsymbol{\sigma}} V^{\dagger}}=\varepsilon\overline{\overline{{\bf z}^{\prime}}{\boldsymbol{\sigma}}}=\varepsilon{\bf z}^{\prime}\overline{\boldsymbol{\sigma}},
\end{equation}
\noindent
where by choosing $V\equiv V_{\bf y}^{\dagger}$ of Eq. (27) we get for  ${\bf z}^{\prime}$
the (30) form.
Finally these manipulations yield for ${\varrho}_{24}$ the canonical form
\begin{equation}
{\varrho}_{24}=\frac{1}{2}\begin{pmatrix}{\kappa}_0+{\kappa}_3&0&0&{\kappa}_1-i{\kappa}_2\\0&\vert\vert{\bf w}\vert\vert^2&-\vert\vert{\bf w}\vert\vert^2&0\\
0&-\vert\vert{\bf w}\vert\vert^2&\vert\vert{\bf w}\vert\vert^2&0\\
{\kappa}_1+i{\kappa}_2&0&0&\kappa_0-\kappa_3\end{pmatrix},
\label{24matrix}
\end{equation}
\noindent
where
\begin{equation}
\kappa_0=\vert\vert{\bf z}^{\prime}\vert\vert^2 =\vert\vert{\bf z}\vert\vert^2,\quad
-\kappa_1=\vert z_1^{\prime}\vert^2-\vert z_2^{\prime}\vert^2,
\quad -\kappa_2=2{\rm Re}(z_1^{\prime}\overline{z_2^{\prime}}),\quad
-\kappa_3=2{\rm Im}(
z_1^{\prime}\overline{z_2^{\prime}}).
\label{hopfmap}
\end{equation}
\noindent
Notice that
\begin{equation}
{\kappa}_0^2={\kappa}_1^2+{\kappa}_2^2+{\kappa}_3^2=\vert\vert{\bf z}\vert\vert^4,
\end{equation}
\noindent
hence the eigenvalues of ${\varrho}_{24}$ are 
$\vert\vert {\bf w}\vert\vert^2,\vert\vert{\bf z}\vert\vert^2,0,0$, i.e. our mixed state is of rank two.
The structure of ${\varrho}_{13}$ is similar with the roles of ${\bf w}$ and ${\bf z}$ exchanged. Following the same steps as in Section III. A. we get for the corresponding squared concurrences the following expressions
\begin{equation}
{\cal C}^2_{13}=\left( \vert\vert {\bf z}\vert\vert^2-\vert{\bf w}^2\vert\right)^2,\qquad
{\cal C}^2_{24}=\left( \vert\vert {\bf w}\vert\vert^2-\vert{\bf z}^2\vert\right)^2
.
\end{equation}
\label{ujconc}
\noindent

Let us now understand the meaning of the invariant $\eta$ from the
four-qubit point of view. It is known that we have four
algebraically independent $SL(2, \compl)^{\otimes 4}$
invariants\cite{Luque, Levay} denoted by $H,L,M$ and $D$. These
are quadratic, quartic, quartic and sextic invariants of the
complex amplitudes ${\Psi}_{ijkl}$ respectively. The invariants
$H$ and $L$ are given by the expressions
\begin{equation}
H=\Psi_0\Psi_{15}-\Psi_1\Psi_{14}-\Psi_2\Psi_{13}+\Psi_3\Psi_{12}-\Psi_4\Psi_{11}
+\Psi_5\Psi_{10}+\Psi_6\Psi_9-\Psi_7\Psi_8, \label{H}
\end{equation}
\noindent and
\begin{equation}
L={\rm Det}\begin{pmatrix}\Psi_0&\Psi_1&\Psi_2&\Psi_3\\
\Psi_4&\Psi_5&\Psi_6&\Psi_7\\
\Psi_8&\Psi_9&\Psi_{10}&\Psi_{11}\\
\Psi_{12}&\Psi_{13}&\Psi_{14}&\Psi_{15}\end{pmatrix}, \label{L}
\end{equation}
\noindent where instead of the binary one we used the decimal
labelling. For the explicit form of the remaining two invariants
$M$ and $D$ see the paper of Luque and Thibon\cite{Luque}. A
straightforward calculation shows that for our four-qubit state we
have $M=D=0$ however,
\begin{equation}
H=-\frac{1}{2}({\bf z}^2+{\bf w}^2),\qquad L=\frac{1}{16}({\bf z}^2-{\bf w}^2)^2,
\label{HL}
\end{equation}
\noindent hence
\begin{equation}\vert L\vert =\frac{1}{16}\eta^2.
\label{ujeta}
\end{equation}
\noindent
For convenience we also introduce the quantity
\begin{equation}
\sigma\equiv\vert{\bf w}^2+{\bf z}^2\vert =2\vert H\vert.
\end{equation}
\noindent
Hence $\eta=\vert{\bf w}^2-{\bf z}^2\vert$ and $\sigma=\vert{\bf w}^2+{\bf z}^2\vert$ are related to the only nonvanishing four qubit invariants $L$ and $H$.
Using the definitions of these quantities and Eq. (13) one can check that
\begin{equation}
{\cal C}^2_{13}=s^2+\frac{1}{2}(\eta^2+\sigma^2)-2\vert\vert{\bf z}^2\vert\vert\vert{\bf w}^2\vert,\quad
{\cal C}^2_{24}=r^2+\frac{1}{2}(\eta^2+\sigma^2)-2\vert\vert{\bf w}^2\vert\vert\vert{\bf z}^2\vert.\quad
\end{equation}
\noindent
Hence we have the inequality
\begin{equation}
{\cal C}^2_{13}+{\cal C}^2_{24}\leq s^2+r^2+{\eta}^2+{\sigma}^2.
\end{equation}
\noindent
Moreover, since ${\cal C}^2_{12}={\cal C}^2_{14}$ after taking the square of Eq.(48) we get
\begin{equation}
{\cal C}^2_{12}+{\cal C}^2_{14}=1-r^2-s^2-\frac{1}{2}{\eta}^2-\sqrt{(1-\eta^2-{\gamma}_-^2)(1-\gamma_+^2)}
\end{equation}
\noindent
Combining this result with Eqs.(64) and (80)
we obtain
\begin{equation}
{\cal C}^2_{12}+{\cal C}^2_{13}+{\cal C}^2_{14}+{\Sigma}_1={\cal C}^2_{1(234)}
\qquad
{\cal C}^2_{12}+{\cal C}^2_{23}+{\cal C}^2_{24}+{\Sigma}_2={\cal C}^2_{2(134)},
\end{equation}
\noindent
where
\begin{equation}
{\Sigma}_1=2\vert\vert{\bf z}\vert\vert^2\vert{\bf w}^2\vert+\sqrt{(\frac{1}{2}\sigma^2+p_+)(\frac{1}{2}\sigma^2+p_-)}-\frac{1}{2}\sigma^2,
\end{equation}
\noindent
\begin{equation}
{\Sigma}_2=2\vert\vert{\bf w}\vert\vert^2\vert{\bf z}^2\vert+\sqrt{(\frac{1}{2}\sigma^2+p_+)(\frac{1}{2}\sigma^2+p_-)}-\frac{1}{2}\sigma^2.
\end{equation}
\noindent
Here
\begin{equation}
p_{\pm}=2\vert\vert{\bf z}\vert\vert^2\vert\vert{\bf w}\vert\vert^2\pm
\frac{1}{2}(4rs-\eta^2).
\end{equation}
\noindent
Notice that by virtue of Eq.(13) $p_-$ is nonnegative.
Moreover according to Eq. (52), for nonseparable states (${\varrho}_{12},\varrho_{14},\varrho_{34},\varrho_{23}$) we have 
nonzero negativity hence $4rs>\eta^2$ hence $p_+$ is also nonnegative.
In this case the {\it residual tangles} $\Sigma_1$ and $\Sigma_2$ as defined by Eqs. (84-85) are positive as they should be, hence the generalized Coffman-Kundu-Wootters inequalities of distributed entanglement\cite{Kundu,Osborne}  
hold
\begin{equation}
{\cal C}^2_{12}+{\cal C}^2_{13}+{\cal C}^2_{14}\leq{\cal C}^2_{1(234)}
\qquad
{\cal C}^2_{12}+{\cal C}^2_{23}+{\cal C}^2_{24}\leq{\cal C}^2_{2(134)}.
\end{equation}
\noindent
For separable states we have ${\cal C}_{12}={\cal C}_{14}={\cal C}_{34}={\cal C}_{23}=0$ and a calculation shows that the (87) inequalities 
in the form ${\cal C}^2_{13}\leq {\cal C}^2_{1(234)}$
and
${\cal C}^2_{24}\leq {\cal C}^2_{2(134)}$
still hold with residual tangles
\begin{equation}
{\Sigma}_1=2\vert\vert{\bf z}\vert\vert^2(\vert{\bf w^2}\vert +\vert\vert{\bf w}\vert\vert^2)\qquad
{\Sigma}_2=2\vert\vert{\bf w}\vert\vert^2(\vert{\bf z^2}\vert +\vert\vert{\bf z}\vert\vert^2).
\end{equation}
\noindent

Eqs. (84), (85) and (88) show the structure of the residual tangle.
Unlike in the well-known three-qubit case these quantities among others contain two invariants $\eta$ and $\sigma$ characterizing four-partite correlations.
The role of $\sigma$  (which for a general four-qubit state is a permutation-invariant)
is to be compared with the similar role
the permutation invariant three-tangle ${\tau}_{123}=4\vert {\cal D}\vert$ plays (an $SL(2,
\compl)^{\otimes 3}$ invariant) within the three-qubit context.
(${\cal D}$ is Cayley's hyperdeterminant\cite{Kundu}.)
An important difference to the three-qubit case is that the residual tangles ${\Sigma}_{1,2}$ seeem to be lacking the important entanglement monotone property.
However, according to a conjecture\cite{haromkinai} the sum ${\Sigma}_1+{\Sigma}_2$ could be an entanglement monotone.
We hope that our explicit form will help to settle this issue at least for
our special four-qubit state of Eqs. (1-2).

\section{Bures metric}
\label{sec_Bur}

As we have emphasized our density matrix ${\varrho}$ can be
regarded as a reduced density matrix of a two-particle system on
$(\compl^2\otimes \compl^2)\wedge(\compl^2\otimes \compl^2)$,
meaning
\begin{equation}
 \varrho = {\Psi}{\Psi}^\dagger,
\end{equation}
where ${\Psi}$ is the $4\times 4$ antisymmetric matrix occurring
in Eq.~(\ref{L}). In the space of such fermionic purifications of
our density matrix the curve ${\Psi}(t)$ is horizontal, when the
differential equation
\begin{equation}
\label{eq_horiz} \dot{\Psi}^\dagger \Psi = \Psi^\dagger \dot{\Psi}
\end{equation}
holds.  We can satisfy this equation by the ansatz
\begin{equation}
 \dot{\Psi}=G\Psi, \qquad G=G^\dagger
\end{equation}
for some $G=G^\dagger$, so that
\begin{equation}
\label{eq_dro}
 d\varrho = \{G,\varrho\}.
\end{equation}
We can define the Bures metric on the space of density matrices,
as follows\cite{Geom}
\begin{equation}
 ds^2_B = \frac{1}{2} \Tr (Gd\varrho).
\end{equation}

Let us now take into account the condition ${\Psi}^T=- \Psi$.
Taking the transpose of  Eq.~(\ref{eq_horiz}), we get
\begin{equation}
\begin{split}
\Psi^T \dot{\Psi}^{\dagger T} &= \dot{\Psi}^T \Psi^{\dagger T},\\
\Psi \dot{\Psi}^\dagger &=  \dot{\Psi} \Psi^\dagger.
\end{split}
\end{equation}
Using this result, we get a simpler formula for $d\varrho$:
\begin{equation}
 d\varrho =  \dot{\Psi} \Psi^\dagger + \Psi \dot{\Psi}^\dagger = 2G\Psi\Psi^\dagger =  2G\varrho.
\end{equation}
and to get $G$, we only have to invert $\varrho$. A calculation of
the eigenvalues of $\varrho$ shows that they are of the
form\cite{LNP}

\begin{equation}
\lambda_{\pm}=\frac{1}{4}(1\pm\sqrt{1-\eta^2}). \label{sajert}
\end{equation}
\noindent Hence ${\varrho}$ is nonsingular iff $\eta\neq 0$ (i.e.
iff $\vert L\vert \neq 0)$. In the following we consider this
case.

For nonsingular density matrices by virtue of 
Eq.~(\ref{eq_Lambda2}), $\varrho^{-1}$ can be calculated easily
\begin{equation}
 \varrho^{-1} = \frac{4}{\eta^2} \left( \idn - \Lambda \right),
\end{equation}
hence:
\begin{equation}
G = \frac{1}{2} d\varrho \varrho^{-1} = \frac{1}{2\eta^2}\left(d\Lambda - d\Lambda \Lambda \right),
\end{equation}
and the Bures-metric:
\begin{equation}
\label{eq_bures}
  ds^2_B = 
  \frac{1}{4\eta^2}\Tr\left(d\Lambda d\Lambda - d\Lambda \Lambda d\Lambda \right).
\end{equation}
Since $\Lambda$ is idempotent and traceless, one can see,
that the trace of the second term equals to zero:
$2\Tr (d\Lambda \Lambda d\Lambda) 
= \Tr (d\Lambda d(\Lambda^2)) =  \Tr (d\Lambda d(-\eta^2) \idn) = 0$.
Let us introduce the quantities $f_{ij} = w_i\cc{z}_j +
\cc{w}_iz_j $. With this notation we have
\begin{equation}
 ds^2_B = \frac{1}{\eta^2}\left( dx_i dx_i +  dy_j dy_j + df_{ij} df_{ij} \right),
\end{equation}
(summation on $i,j=1,2,3$ is implied.) Moreover, a calculation
shows that $\eta^2 = 1-(x_i x_i +  y_j y_j + f_{ij} f_{ij})$, so
after putting the  quantities $x_i$, $y_j$, $f_{ij}$ into a $15$
component vector  $\ve{k}\in\real^{15}$ our final result is the
nice formula
\begin{equation}
  ds^2_B = \frac{1}{1-\ve{k}^2}d\ve{k}^2 = \frac{1}{4} \eta^2 \left[ \frac{4d\ve{k}^2}{(1-\ve{k}^2)^2} \right].
  \label{Bures}
\end{equation}

Let us compare this formula with the one obtained for the Bures
metric of one-qubit density matrices arising as a reduced density
matrix from a pair of distinguishable qubits\cite{Levay2}
\begin{equation}
dl^2_B=\frac{1}{ 4\cosh^2\beta  }\left(d\beta^2+\sinh^2\beta
 d{\Omega}^2\right), \label{quat}
\end{equation}
\noindent where $1-C^2=\tanh^2\beta$ with $C$ is the {\it pure
state concurrence} for two qubits, and $d\Omega^2$ is the usual
line element on the two-sphere $S^2$ expressed by the angular
coordinates $\vartheta$ and $\varphi$. Since the space of
one-qubit density matrices is the Bloch-ball ${\bf B}^3$ this
parametrization provides a map between the upper sheet of the
double sheeted hyperboloid ${\bf H}^3$ and ${\bf B}^3$. The
standard metric on ${\bf H}^3$ is just $d{\beta}^2+\sinh^2\beta
d{\Omega}^2$. Hence we see that the Bures metric is up to the
conformal factor $C^2/4$ is just the standard metric on the upper
sheet of the double sheeted hyperboloid $H^3$. However, using the
stereographic projection one can show that
\begin{equation}
d{\beta}^2+\sinh^2\beta d{\Omega}^2=\frac{4{\bf dR}^2}{(1-{\bf
R}^2)^2}, \label{stereo}
\end{equation}
\noindent where $R_1R_2$ and $R_3$ can alternatively be used to
parametrize ${\bf B}^3$. Hence we can write
\begin{equation}
dl^2_B=\frac{1}{4}C^2\left[\frac{4{\bf dR}^2}{(1-{\bf
R}^2)^2}\right] \label{szepalak}
\end{equation}
\noindent where the metric on the right is the standard Poincar\'e
metric on the unit ball which is now just the Bloch-ball.
Comparing this equation with our previous expression of 
Eq.~(\ref{Bures}) we see that it is up to the conformal factor
$\eta^2/4$ is just the Poincar\'e metric on the Poincar\'e ball
${\bf B}^{15}$.
We emphasize however, that unlike the usual one-qubit mixed state where all the Bloch parameters characterizing the density matrix are independent, here the $15$ parameters associated to the vector ${\bf k}$ are subject to nontrivial constraints.
These constraints describe some nontrivial embedding of the space of nonsingular density matrices ${
\cal D}$ into the Bloch ball ${\bf B}^{15}$ with our Bures metric of Eq. (101).

\section{Conclusions}
\label{sec_Con} In this paper we investigated the structure of a
$12$ parameter family of two-qubit density matrices with fermionic
purifications. Our starting point was a four-qubit state with a
special antisymmetry constraint imposed on its amplitudes. Such
states are elements of the space 
$(\compl^2\otimes \compl^2)\wedge(\compl^2\otimes \compl^2)$ and the admissible local
operations are of the form $(U\otimes V)\otimes (U\otimes V)\in
SL(2, \compl)^{\otimes 4} $. Our density matrices are arising as
the reduced ones ${\varrho}={\varrho}_{12}={\rm
Tr}_{34}(\vert\Psi\rangle\langle\Psi\vert)$. Since the $12$
subsystem is indistinguishable from the $34$ one we have
${\varrho}_{12}={\varrho}_{34}$. We obtained an explicit form for
${\varrho}$ in terms of the $6$ independent complex amplitudes
${\bf w}$ and ${\bf z}$ of our four-qubit states. Employing local
unitary transformations of the form $U\otimes V\in SU(2)\times
SU(2)\subset SL(2, \compl)\times SL(2, \compl)$ we derived the
canonical form for ${\varrho}$. This form enabled an explicit
calculation for different entanglement measures. We have
calculated the Wootters concurrence, the negativity, and the
purity. The quantities occurring in these formulae (and some additional ones) are subject to monogamy relations of distributed entanglement similar to
the ones showing up in the Coffman-Kundu-Wootters relations for
three-qubits. They are  characterizing the entanglement trade off between
different subsystems. We have entanglement measures $\eta$ and $\sigma$
 describing the intrinsically four-partite
correlations,  quantities ( Wootters concurrences) keeping track
the mixed state entanglement of the bipartite subsystems embedded in the
four-qubit one. Finally we have the independent quantities ${\cal
C}^2_{1(234)}$ and ${\cal C}^2_{2(134)}$ measuring how much
subsystems $1$ and $2$ are entangled {\it individually} to the
rest. We derived explicit formulas displaying how these important
quantities are related. At last we have studied in some detail the
Bures geometry underlying the structure of these density matrices.
We have shown that the constraint of antisymmetry makes it
possible to obtain a nice explicit formula for the Bures metric reminiscent of the ones known from hyperbolic geometry\cite{Ungar}.

\section{Acknowledgment}
\label{sec_Ack} One of us (P. L.) would like to express his
gratitude to Professor Werner Scheid for his warm hospitality at
the Department of Theoretical Physics of the Justus Liebig
University of Giessen where part of this work was completed.
Financial support from the Orsz\'agos Tudom\'anyos Kutat\'asi Alap (OTKA)
(Grants No. T046868, T047041, and T038191) is gratefully acknowledged.

\appendix
\section{Upper bound of negativity}
\label{sec_app}
In this fermionic-correlated case, defined by equations
(\ref{eq_rho}), (\ref{eq_Lambda}), (\ref{eq_xy}) and (\ref{eq_ceta}),
we can prove the following inequality:

\textit{Theorem:}
For all entangled $\varrho$:
\begin{equation}
\label{eq_upperb}
 \negat(\varrho) \leq \frac{1}{2} \left( \sqrt{ 2 - (1-2\conc(\varrho))^2 } -1 \right).
\end{equation}

\textit{Proof:}
Insert Eqs.~(\ref{eq_conc}) and (\ref{eq_negat}) into (\ref{eq_upperb}):
\begin{equation}
\frac{1}{2} \left( \sqrt{ 1 - \eta^2 + 4rs } -1 \right) 
\leq \frac{1}{2} \left( \sqrt{ 2  -  
\left(1- \sqrt{1-\gamma_-^2-\eta^2}+ \sqrt{1-\gamma_+^2}\right)^2 } -1 \right),
\end{equation}
after some algebra, we can rearrange the terms:
\begin{equation}
\begin{split}
0
&\leq
2\eta^2 -2 +\gamma_-^2 + \gamma_+^2 - 4rs \\
& \qquad  +2\sqrt{1-\gamma_-^2-\eta^2} 
-2\sqrt{1-\gamma_+^2}
+2\sqrt{1-\gamma_-^2-\eta^2}\sqrt{1-\gamma_+^2},
\end{split}
\end{equation}
If follows from the definition (\ref{gammapm})
that $\gamma_+^2- 4rs = \gamma_-^2 $. With this:
\begin{equation}
0 \leq
-(1-\gamma_-^2-\eta^2)
+\sqrt{1-\gamma_-^2-\eta^2} 
-\sqrt{1-\gamma_+^2}
+\sqrt{1-\gamma_-^2-\eta^2}\sqrt{1-\gamma_+^2}.
\end{equation}
The right-hand side is factorizable:
\begin{equation}
0 \leq
\left( \sqrt{1-\gamma_-^2-\eta^2} - \sqrt{1-\gamma_+^2 } \right)
\left( 1 - \sqrt{1-\gamma_-^2-\eta^2} \right).
\end{equation}
The second parenthesis is obviously positive.
For entangled states  $\conc(\varrho)>0$,
and the first parenthesis is proportional to the concurrence,
which is strictly positive.
\textit{Q.E.D.}


\begin{thebibliography}{}
\bibitem{Nielsen} M. A. Nielsen and I. L. Chuang, {\it Quantum
Computation and Quantum Information}, Cambridge University Press,
Cambridge, England, 2000.
\bibitem{Geom} Ingemar Bengtsson, Karol \.Zyczkowski,
{\it Geometry of Quantum States}, Cambridge University Press, 2006.
\bibitem{Petz} M. Ohya and D. Petz, {\it Quantum entropy and its
use}, Springer Verlag, Berlin, 1993.
\bibitem{Hill}W. K. Wootters, Phys. Rev. Lett. {\bf 80}, 2245
(1998), S. Hill and W. K. Wootters, Phys. Rev. Lett. {\bf 78}, 5022
(1997).
\bibitem{hivatbengtsson} For a nice summary see the relevant
chapter of Ref.2
\bibitem{Uhlmann} A. Uhlmann, Phys. Rev. {\bf A63} 032307 (2000).
\bibitem{Siewert} A. Osterloh and J. Siewert, Phys. Rev. {\bf
A72}, 012337 (2005).
\bibitem{Fuchs} C. M. Caves, C. A. Fuchs and P. Rungta, Found.
Phys. Lett. {\bf 14}, 199 (2001).
\bibitem{Verstraete}F. Verstraete, K. Audenaert, J. Dehaene, B. De
Moor, J. Phys. {\bf A34}, 10327 (2001). K. Audenaert, F.
Verstraete, T. De Bie, B. De Moor, arXiv:quant-ph/0012074.
\bibitem{Ishizaka} S. Ishizaka and T. Hiroshima, Phys. Rev. {\bf
A62}, 022310 (2000).
\bibitem{Eisert} J. Eisert and M. B. Plenio, J. Mod. Opt. {\bf 46},
145 (1999).
\bibitem{Kundu} V. Coffman, J. Kundu and W. K. Wootters, Phys.
Rev. {\bf A61}, 052306 (2000).
\bibitem{Zanardi} P. Zanardi, D. Lidar and S. Lloyd, Phys. Rev.
Lett. {\bf 92}, 060402 (2004).
\bibitem{Osborne} T. Osborne, F. Verstraete,
Phys. Rev. Lett. {\bf 96}, 220503 (2006).
\bibitem{Schlie} J. Schliemann, J. I. Cirac, M. Kus, M. Lewenstein
and D. Loss, Phys. Rev. {\bf A64}, 022303 (2001).
\bibitem{LNP}
 P. L\'evay, Sz. Nagy, J. Pipek, Phys. Rev. {\bf A72}, 022302 (2005).
\bibitem{Luque} J-G. Luque and J-Y. Thibon, Phys. Rev. {\bf A67}
042303 (2003).
\bibitem{Peres} A. Peres, Phys. Rev. Lett. {\bf 77}, 1413 (1996).
\bibitem{Levay} P. L\'evay, J. Phys. {\bf A39}, 9533 (2006).
\bibitem{Levay2} P. L\'evay, J. Phys. {\bf A37}, 1821 (2004).
\bibitem{haromkinai} Y-K Bai, D. Yang, Z. D. Wang,
Phys. Rev. {\bf A76}, 022336 (2007).
\bibitem{Ungar} J. Chen, L. Fu, A. A. Ungar and X. Zhao, Phys. Rev. {\bf A65}
024303 (2002).
\end{thebibliography}
\end{document}